\documentclass[twocolumn,prl,preprint,superscriptaddress,showpacs,showkeys]{revtex4}

\usepackage{graphicx}
\usepackage{amsmath}
\usepackage{amssymb}
\usepackage{amsfonts}

\begin{document}

\title{Inverse lift: a signature of the elasticity of complex fluids?}

\author{Benjamin Dollet}
\email{bdollet@spectro.ujf-grenoble.fr} \affiliation{Laboratoire de
Spectrom\'etrie Physique, BP 87, 38402 Saint-Martin-d'H\`eres Cedex, France}
\altaffiliation{UMR 5588 CNRS and Universit\'e Joseph Fourier.}
\author{Miguel Aubouy}
\affiliation{SI3M, DRFMC, CEA, 38054 Grenoble Cedex 9, France}
\author{Fran\c cois Graner}
  \affiliation{Laboratoire de
Spectrom\'etrie Physique, BP 87, 38402 Saint-Martin-d'H\`eres Cedex, France}
\altaffiliation{UMR 5588 CNRS and Universit\'e Joseph Fourier}

\date{\today}

\begin{abstract}
To understand the mechanics of a complex fluid such as a foam we propose a
model experiment (a bidimensional flow around an obstacle) for which an
external sollicitation is applied, and a local response is measured,
simultaneously. We observe that an asymmetric obstacle (cambered airfoil
profile) experiences a downards lift, opposite to the lift usually known (in a
different context) in aerodynamics. Correlations of velocity, deformations and
pressure fields yield a clear explanation of this inverse lift, involving the
elasticity of the foam. We argue that such an inverse lift is likely common to
complex fluids with elasticity.
\end{abstract}



\maketitle

A liquid foam exhibits   ``complex" behaviour under stress:  it is elastic for
small deformation, plastic for large deformation, and flows at large
deformation rates \cite{Weaire1999,jiangpre,Saint-Jalmes1999}. This rich
mechanical behavior is used in many of the foams' applications, including ore
separation by flotation in mines, drilling and extraction in oil industry, and
cleaning in confined media such as pipes  \cite{Weaire1999}. A foam is a
convenient model to study constitutive relations, since the microscale is the
scale of bubbles (not of molecules, as in most complex fluids,  such as
emulsions \cite{Mason1995,Mason1996}, colloids and polymer solutions
\cite{Phan-Thien2002,Macosko1994,Tanner2000,Bird1987}), and is easily
observable. In particular, a foam with only one bubble layer (so-called
``two-dimensional foam'' \cite{Weaire1999,cox2003})  is easy to image, and
image analysis yields information on all the  geometrical properties of the
foam.

We perform a Stokes experiment
\cite{deBruyn2004,Dollet2005a,Alonso2000,Asipauskas2003,Mitsoulis2004,Roquet2003},
i.e. we study the flow of foam around obstacles, using a set-up fully described
in Ref. \cite{Dollet2005a}. Briefly, a tank is filled with a bulk solution
obtained by adding 1\% of commercial dish-washing liquid (Taci, Henkel) to
desionised water. Its surface tension, measured with the oscillating bubble
method, is $\gamma = 26.1\pm 0.2$ mN m$^{-1}$, and its kinematic viscosity,
measured with a capillary viscosimeter, is $1.06\pm 0.04$ mm s$^{-2}$. Nitrogen
is blown in the solution through a nozzle or a tube at a computer controlled
flow rate. This generates a foam, constituted by a horizontal monolayer of
bubbles of average thickness $h_0 = 3.5$ mm, confined between the bulk solution
and a glass top plate \cite{cox2003}. The foam is monodisperse (bubble area at
channel entrance: $A_0 = 16.0\pm 0.5$ mm$^2$) and its fluid fraction is around
10\% (the evaluation of this quantity in such a setup will be detailed in
future work). It flows around an obstacle placed at the middle of the channel.
The obstacle is linked to a fixed base through an elastic fiber; we thus
measure the force exerted by the flowing foam on the obstacle (precision $<
0.1$ mN) by tracking the obstacle displacement from its position at rest, using
a CCD camera which images the foam flow from above. The flow rate is 50 ml
min$^{-1}$, and the average velocity 2.7 mm s$^{-1}$, except in Fig.
\ref{Forces}.

\begin{figure}
\includegraphics[width=8cm]{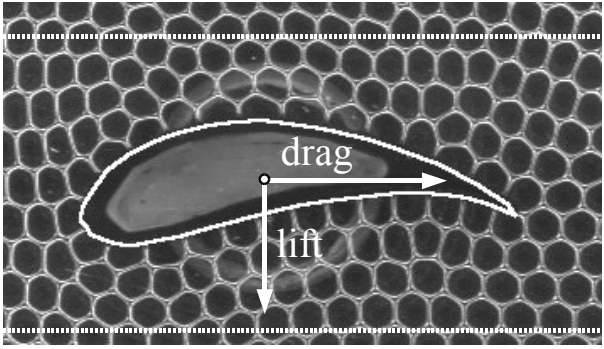}
\caption{\label{Image} Top view of the airfoil with a small part of the flowing
foam. The superimposed white line indicates its boundary. It is a Joukovski
profile \cite{Huerre1998}, which equation of shape writes $x(t) = \{ 1.56 +
[(0.168 + \cos t)^2 + (0.25 + \sin t)^2]^{-1} \} (0.168 + \cos t)$ and $y(t) =
\{ 1.56 - [(0.168 + \cos t)^2 + (0.25 + \sin t)^2]^{-1} \} (0.25 + \sin t)$,
with lengths in centimeters, the angle $t$ ranging from $-\pi$ to $\pi$. Its
length from leading to trailing edge is 5.2 cm. The arrows indicate the
direction of drag and lift (flow from left to right), and the white dot marks
the contact point between the airfoil and the fiber. The dotted lines show
where the quantities displayed in Fig. \ref{HautBas} are evaluated. A movie is
available at \texttt{www-lsp/link/mousses-films.htm}.}
\end{figure}

\begin{figure}
\includegraphics[width=8cm]{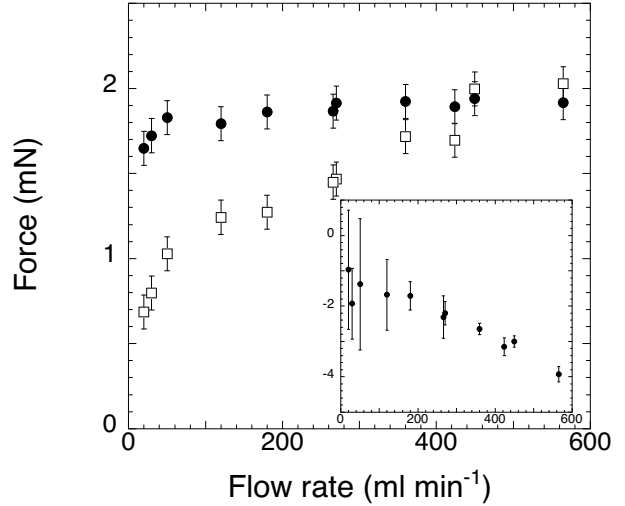}
\caption{\label{Forces} Forces exerted by the flowing foam on the airfoil
\emph{versus} flow rate: drag ($\square$) and lift ($\bullet$). {\it Insert}
Leading angle ($^\circ$) spontaneously selected by the airfoil \emph{versus}
flow rate (ml min$^{-1}$); the standard deviation of its time fluctuations are
plotted as error bars.}
\end{figure}

Here, the obstacle is a cambered airfoil (Fig. \ref{Image}). Like every
obstacle in a flow, the airfoil experiences a streamwise force, the drag; but
owing to its asymmetry, it also feels a torque, and a spanwise force: the lift.
The obstacle  is free to rotate around the contact point with the fiber. We
quantify its (zero-torque) stable equilibrium  orientation by measuring the
leading angle $\alpha$, defined as the angle between the axis passing through
the points $x=\pm 1$, $y=0$ in the Joukovski equation (see caption of Fig.
\ref{Image}), and the flow direction. The leading angle (which depends on the
location of the contact point,  and hence is not generic) is small and
negative: it decreases from $-1^\circ$ to $-4^\circ$ in the studied range of
flow rate (Fig. \ref{Forces}).

Fig. \ref{Forces} reports the zero-torque orientation, and corresponding drag
and lift measurements,  versus the flow rate. It evidences a non-zero drag at
vanishing flow rate, which is the force required to trigger a steady motion of
the foam with respect to the obstacle, and appears more as a solid-like
property. On the other hand, the drag is an increasing affine function of the
flow rate, as expected \cite{Dollet2005a,Alonso2000} (and its value is almost
as low as that of a non-cambered airfoil \cite{Dollet2005a}): this is a
consequence of the fluid-like properties of the foam \cite{Dollet2005a}.

Fig. \ref{Forces} also shows a striking feature: the lift is directed
downwards. This is opposite to the lift which appears (in an entirely different
physical regime) in aerodynamics \cite{Huerre1998}.
  To
our knowledge, this is the first time that such an inverse lift is
experimentally evidenced. Does it originate from solid- or liquid-like
property?

\begin{figure}
\includegraphics[width=8cm]{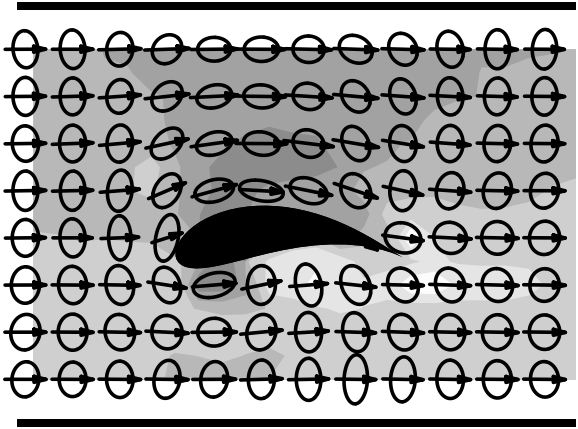}
\caption{\label{VMA} Experimental measurements of velocity, area and
deformation fields around the airfoil across the whole channel (width: 10 cm).
Each measurement results from an average over a representative volume element
(square box of side 1.1 cm) and over time (750 successive images in steady
regime). Arrows: velocity of bubble centers of mass. Background gray levels:
bubble area, with 12\% relative variation between the most (dark grey) and less
(white) compressed bubbles: the mean area  corresponds to the contour line
between the two gray levels at the left side of the figure. Ellipses: texture
tensor, the major axis representing the direction and magnitude of maximal
bubble elongation (an isotropic region would be represented by a circle).}
\end{figure}

\begin{figure}
\includegraphics[width=8cm]{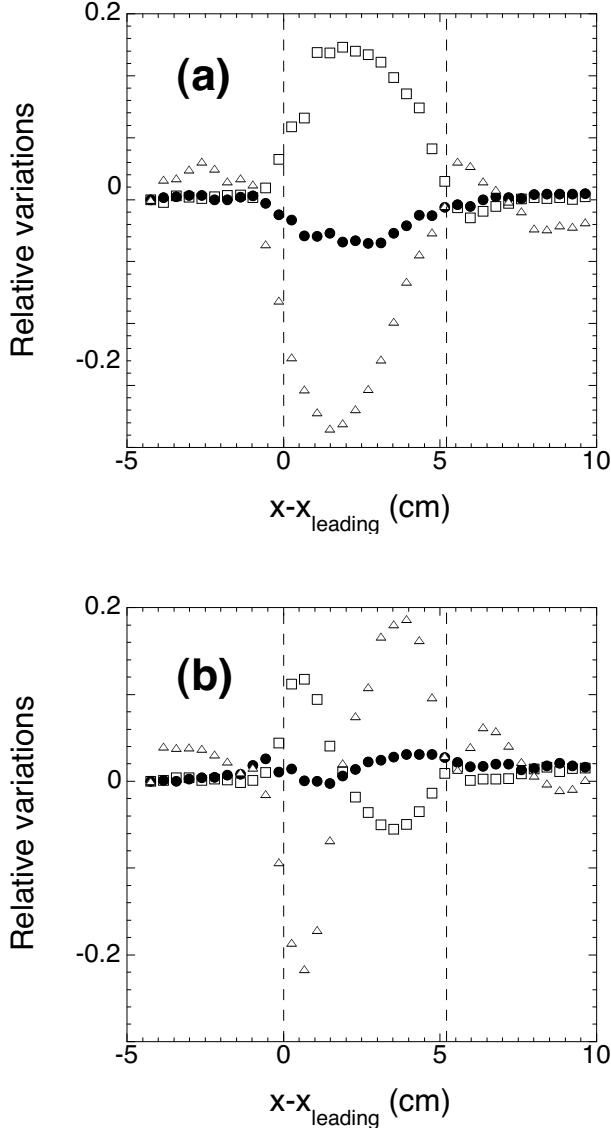}
\caption{\label{HautBas} Experimental measurements of velocity, area and
deformation fields (a) 1 cm above and (b) 1 cm below the airfoil. Same data as
Fig. \ref{VMA}. The  streamwise component $v_x$ of the velocity ($\square$),
the bubble area ($\bullet$) and the
  $yy$ component of
the texture tensor, $M_{yy}$ ($\triangle$) are adimensioned by their value at
the channel entrance (2.7 mm$\cdot$s$^{-1}$, 16 mm$^2$ and 3.5 mm$^2$,
respectively), and we represent their variations relative to these values
\emph{versus} the streamwise coordinate $x$ relative to the leading edge.
Vertical dots indicate the leading and trailing edges. The inversion below the
airfoil occurs 1.8 cm after the leading adge.}
\end{figure}

As a first hint, we note that  the lift hardly increases with the flow rate. To
understand its physical origin, we now turn to the effect of the obstacle on
the foam flow. Using bubbles as passive tracers, we perform local measurements
of the velocity, area and deformation fields (which correlate respectively with
viscous, surface tension and pressure contributions to the stress
\cite{Asipauskas2003,Aubouy2003,Janiaud2005}). Their time averages are plotted
around the airfoil (Fig. \ref{VMA}), and along two horizontal lines, 1 cm above
and 1 cm below the airfoil (Fig. \ref{HautBas}).

The velocity field shows that the convex regions of the airfoil constrict the
flow. At the trailing edge's cusp, the velocity field is regular. It
does not exhibit
singularity, nor any qualitative difference with aerodynamics, where
the  velocity at a sharp
trailing edge is continuous (Kutta condition
\cite{Batchelor2000,Huerre1998}).

The 3D compressibility of bubble gas is generally neglected (foams compression
modulus, of order of atmospheric pressure, is typically three orders of
magnitude larger than their shear modulus \cite{Weaire1999}). But here, thanks
to the bulk solution in contact with the bottom of the foam, an increase of
pressure increases the height of the bubbles (which equalise their pressure
with the hydrostatic pressure of the bulk solution). Its effect is a decrease
of the visible bubble area. The present foam thus has an effective 2D
compressibility, equal to $(\rho gh_0)^{-1} = 2,9\times 10^{-2}$ Pa$^{-1}$
\cite{Dollet2005a}. For such a compressibility, bubble area variations act as a
passive tracer of the pressure field: they are large enough to be measurable,
and small enough not to perturb the flow. From area measurements, we can
determine the net contribution of pressure to the force: it writes
\cite{Dollet2005a} $\vec{F}_P = -\rho gA_0^2 h_0^2 \oint \mathrm{d}\ell
\,\vec{n}/A^2$, where $\rho = 1.00\times 10^3$ kg m$^{-3}$ is the volumetric
mass of the solution, $g=9.8$ m s$^{-2}$ the gravity acceleration, $h_0$ and
$A_0$ the average values of the bubbles' depth and area, the integral being
taken over the contour of the airfoil ($\vec{n}$ is the outwards normal of the
contour and $\mathrm{d}\ell$ its length element). We have measured $\vec{F}_P$.
It contributes for 0.20 mN to the drag, and for 0.74 mN to the downwards lift.
It is worth noting that the bubbles' pressure increases where the flow
accelerates (Figs. \ref{VMA} and \ref{HautBas}), contrary to Newtonian fluids
in inertial flow \cite{Batchelor2000}: this is a clear signature of foam
elasticity.

We quantify the deformations of bubbles, visible on an image
\cite{Asipauskas2003}, by measuring the (always symmetric) texture tensor
\cite{Aubouy2003,Janiaud2005}, defined as: $\bar{\bar{M}} = \langle \vec{\ell}
\otimes \vec{\ell} \rangle = ( \langle  {\ell}_x^2 \rangle, \langle
{\ell}_x{\ell}_y \rangle, \langle  {\ell}_x{\ell}_y \rangle, \langle
{\ell}_y^2\rangle )$. It only requires to measure the bubble edge vectors
$\vec{\ell}$ linking two neighbouring vertices; the average is taken over a
representative volume element. Fig. \ref{VMA} shows the elongation of the
bubbles on both sides of the airfoil: vertical stretching in the concave region
of the airfoil (below the trailing edge); horizontal stretching in the convex
regions (above the airfoil and below the leading edge). This qualitatively
different behaviour appears clearly on the $M_{yy}$ plots (Fig. \ref{HautBas}).

Correlations visible on Fig. \ref{VMA} and \ref{HautBas} yield a physical
explanation of the downwards lift. In convex regions, above the airfoil or
below at $x<1.8$ cm, the flow is constricted and accelerates, and bubbles
stretch streamwise. Since the elastic stress in foams is due to surface tension
and correlated to the orientation of bubble edges
\cite{Weaire1999,Reinelt2000,Asipauskas2003,Janiaud2005}, the direction of main
elastic stress is streamwise. In the concave region, below the airfoil at
$x>1.8$ cm, the picture is reversed, and the direction of main elastic stress
is spanwise; this contributes to a downwards lift, like the bubbles' pressure
contribution described above. The net balance is a resulting downwards lift.
This explanation is in principle only valid in a quasistatic regime, but since
the lift does not increase significantly with the flow rate (Fig. 2), it seems
to constitute the essential ingredient in the studied range of flow rate.
Moreover, the ``T1" bubble neighbour-swappings \cite{Weaire1999} saturate the
maximum value of deformation but do not fundamentally affect this mechanism.
The lift thus appears mainly as an elastic effect, typical of a solid-like
behaviour.

To estimate the elastic contribution to the lift, we approximate the foam by a
2D one, with a line tension $\lambda = 2\gamma h=0.18$ mN, probably a strong
underestimation (due to 3D effects, the effective height and thus line tension
is probably larger). The 2D elastic stress in such a 2D foam would write
\cite{Weaire1999,Janiaud2005} as: $\bar{\bar{\sigma}}_{\mathrm{el}} =
\lambda\rho \langle \vec{\ell} \otimes \vec{\ell}/\ell \rangle$, where
$\vec{\ell}$ denotes a bubble edge,  and $\rho = 3/A$ the areal density of
edges
  (there are in average six edges per bubble \cite{Weaire1999}, each one being shared by two bubbles).
We estimate
  an elastic contribution of 0.05 mN to the drag, and 0.23
mN to the lift.
   Adding pressure and elastic contributions indicates
values of 0.25 mN for the drag and 0.97 mN for the downwards lift, of the
same order but lower that the measured forces.

At this stage, we can propose an explanation for the
low dependence of the lift on the flow rate. Since the foam slips
along the airfoil, in the
lubrication films between the airfoil and the surrounding bubbles there appears
  strong velocity gradients
\cite{Cantat2004}; they are  perpendicular to the airfoil boundary, hence
mainly perpendicular to the flow. The resultant of
the viscous friction thus contributes much more to the drag than to
the lift.

To discuss whether this lift is a true effect of foam physics, we must examine
possible other contributions. First, to exclude possible artifacts linked with
the present set-up, namely bubble 3D geometry and their effective 2D
compressibility, S. J. Cox (private communication) performed simulations of a
true 2D, incompressible foam flow using the Surface Evolver software
\cite{Brakke1992}. He unambiguously observed the same bubble stretching and
downwards lift, due both to the bubble edges' surface tension and to the
pressure contribution. Second, the confinement of the foam by the sides of the
channel is expected to play a role: this is always the case in 2D flows around
obstacles, either Newtonian \cite{Faxen1946} or non-Newtonian
\cite{Mitsoulis2004,Roquet2003}. However, the relevant parameter is the
logarithm of the channel width to obstacle size ratio, and experimental studies
of the drag exerted by a flowing foam on obstacles show weak variations of the
drag with the ratio obstacle size/channel width \cite{deBruyn2004,Dollet2005a}.
Hence, we expect a weak quantitative (and no qualitative) effect of the channel
width on the lift. Third, the aerodynamic lift scales like $U\sin
(\alpha+\beta)$ \cite{Batchelor2000,Huerre1998}, where $U$ denotes the relative
velocity of the flow and the obstacle, and $\beta$ the purely geometric camber
angle. For our airfoil this angle equals $\beta = 14.5^\circ$; hence, even if
$\alpha$ is negative,  $\alpha+\beta$ remains positive, and this cannot explain
our observations. Fourth, there is an average pressure gradient $\nabla P$ due
to the dissipation of flowing foam \cite{Cantat2004,Dollet2005b}. It equals 40
Pa m$^{-1}$ for a flow rate of 50 ml min$^{-1}$, and reaches $1.7\times 10^2$
Pa m$^{-1}$ at the highest studied flow rate (565 ml min$^{-1}$, see Fig.
\ref{Forces}). It
  slightly prestrains the
bubbles before arriving on the obstacle, but this does not qualitatively affect
the main features of the deformation. It also adds an Archimedes thrust-like
downstream contribution to the drag: $\Pi = Sh_0 \nabla P$ ($S=7.74$ mm$^2$:
surface of the airfoil), which is not negligible ($\nabla P = 0.11$ mN at 50 ml
min$^{-1}$ and up to 0.46 mN at 565 ml min$^{-1}$), but which does not
contribute to the lift.

How generic is the inverse lift? First, it is compatible with other phenomena
(for instance, die swell, Weissenberg rod-climbing effect
\cite{Phan-Thien2002,Macosko1994,Tanner2000,Bird1987}, sedimentation of
particles \cite{Huang1998}, or inverse Magnus effect \cite{Wang2004}) observed
or predicted with non-Newtonian fluids which act in the opposite sense to
Newtonian fluids in inertial flow \cite{Batchelor2000}.  More precisely,
wherever the pressure of a Newtonian fluid would push an obstacle, the normal
stress in a viscoelastic fluid pulls it; for instance it can change from
compression to tension at the trailing edge \cite{Wang2004}, in agreement with
what we observe here. Second, preliminary studies of the flow of a second-order
fluid on the same airfoil profile show unambiguously the inverse lift
\cite{Wang2005}. Note however that (contrary to the present case) for fluids
with zero yield stress, the lift would be expected to vanish at vanishing flow
rate, and it may increase significantly with the flow rate if normal stress
differences do \cite{Wang2005}. Third, we experimentally let an asymmetric
object (a truncated portion of a disk, with a circular side and a straight one)
settle under gravity in a model viscoelastic fluid (0.5\% w:w cellulose
solution) confined between vertical plates of glass. The object does  feel a
lift directed from the most to the less convex side. Several arguments thus
suggest that such an inverse lift is expected to be generic to other fluids
which can store elasticity.

\begin{acknowledgements}
We warmly thank S. J. Cox  for courteously performing simulations, F. Elias, C.
Quilliet and C. Raufaste for experimental help, R. Gros for fabrication of the
airfoil, A. Saint-Jalmes, S. Marze, and A. Viallat for measurements
of the solution properties, and T. Podgorski, C. Verdier and J. Wang
for enlightening
discussions.
\end{acknowledgements}

\end{document}